\documentclass[10pt,a4paper]{article}
\usepackage[left=2cm,right=2cm,top=2cm,bottom=2cm]{geometry}
\usepackage[english]{babel}
\usepackage{amsmath}
\usepackage{amsfonts}
\usepackage{amssymb}
\usepackage{bbm}
\usepackage{latexsym}
\usepackage{lmodern}
\usepackage{mathrsfs}
\usepackage{slashed}
\usepackage{braket}
\usepackage{tensor}
\usepackage{simplewick}
\usepackage{graphicx}
\usepackage{xcolor}
\usepackage[font=small,labelfont=it]{caption}
\usepackage[font=scriptsize]{caption}
\usepackage[hidelinks]{hyperref}
\hypersetup{
        colorlinks=true,
        linkcolor=blue,
        citecolor=magenta,
}
\usepackage[nottoc]{tocbibind} % Puts references in toc

\def\C{\mathbb C}

\renewcommand{\>}{\rangle}
\newcommand{\<}{\langle}

\newcommand{\sfrac}[2]{{\textstyle\frac{#1}{#2}}}

\newcommand{\pa}{\partial}
\newcommand{\im}{{\mathrm{i}}}

\newcommand{\ep}{{\mathrm{e}}}
\newcommand{\diff}{{\mathrm{d}}}

\newcommand{\beq}{\begin{equation}}
\newcommand{\eeq}{\end{equation}}
\newcommand{\eq}{\end{equation}}
\newcommand{\bea}{\begin{eqnarray}}
\newcommand{\eea}{\end{eqnarray}}

\renewcommand{\and}{{\quad{\rm and}\quad}}
\newcommand{\und}{{\qquad{\rm and}\qquad}}
\renewcommand{\=}{\ =\ }

\allowdisplaybreaks  % allow to break the content of align between page
\DeclareMathOperator{\tr}{tr}

\newcommand{\ie}{i.e.\ }
\newcommand{\eg}{e.g.\ }
\newcommand{\GroupName}[1]{ { \text{#1} } }

\newcommand{\Lagr}{ { \mathcal{L} } }
\newcommand{\Delt}{ { \mathcal{M} } }

\advance \textheight by \topskip
\makeatletter
\@addtoreset{equation}{section}

\makeatother
\usepackage{tikz}
\usetikzlibrary{positioning,arrows}
\usetikzlibrary{decorations.pathmorphing}
\usetikzlibrary{decorations.markings}
\tikzset{
particle/.style={thick,draw=black, postaction={decorate},
    decoration={markings,mark=at position .50 with {\arrow[line width=1pt]{>}}  }},
aparticle/.style={thick,draw=black, postaction={decorate},
    decoration={markings,mark=at position .5 with  {\arrow[line width=1pt]{<}}  }},
gluon/.style={decorate, draw=black,
    decoration={coil,aspect=0}}
 }
\begin{document}

\begin{titlepage}
\setcounter{page}{0}

\phantom{.}
\vskip 1.5cm

\begin{center}

{\LARGE \bf 
Nicolai maps with four-fermion interactions
}
\vspace{12mm}

{\Large Lorenzo Casarin$^{a,b}$, Olaf Lechtenfeld$^{a}$ \ and \ Maximilian Rupprecht$^{a}$}
\\[8mm]
\noindent ${}^a${\em
Institut f\"ur Theoretische Physik and Riemann Center for Geometry and Physics\\
Leibniz Universit\"at Hannover, Appelstrasse 2, 30167 Hannover, Germany}
\\[6mm]
\noindent ${}^b${\em
Max-Planck-Institut f\"ur Gravitationsphysik (Albert-Einstein-Institut)\\
Am M\"uhlenberg 1, 14476 Potsdam, Germany}
\vspace{12mm}

\begin{abstract} 
\noindent
Nicolai maps offer an alternative description of supersymmetric theories
via nonlinear and nonlocal transformations  characterized by 
the so-called `free-action' and `determinant-matching' conditions. 
The latter expresses the equality of the Jacobian determinant of the transformation 
with the one obtained by integrating out the fermions, 
which so far have been considered only to quadratic terms. 
We argue that such a restriction is not substantial, as Nicolai maps can be constructed 
for arbitrary nonlinear sigma models, which feature four-fermion interactions. 
The fermionic effective one-loop action then gets generalized to higher loops 
and the perturbative tree expansion of such Nicolai maps
receives quantum corrections in the form of fermion loop decorations.
The `free-action condition' continues to hold for the classical map,
but the `determinant-matching condition' is extended to an infinite hierarchy
in fermion loop order.
After general considerations for sigma models in four dimensions, 
we specialize to the case of \(\mathbb C \mathrm P ^N\) symmetric spaces 
and construct the associated Nicolai map. These sigma models admit a formulation 
with only quadratic fermions via an auxiliary vector field, 
which  does not simplify our analysis.
\end{abstract}

\end{center}

\end{titlepage}

\tableofcontents

\section{Introduction and summary}

\noindent
The Nicolai map~$T$~\cite{Nic1,Nic2,Nic3} 
is a (generically nonlocal and nonlinear) field transformation that relates 
supersymmetric theories at different values of their parameters, say coupling constants~$g$.
In particular, it allows one to compute the expectation value of any operator~$Y$ 
built from the bosonic fields~$\phi$ at coupling~$g$ by evaluating a free-field ($g{=}0$)
correlator of the inversely transformed operator~$T^{-1}Y$,\footnote{
The vanishing of the vacuum energy in supersymmetric theories properly normalizes $\langle1\rangle_g=1$.}
\begin{equation} \label{Tdef}
\bigl\langle Y[\phi] \bigr\rangle_g 
\= \bigl\langle (T_g^{-1}Y)[\phi] \bigr\rangle_0
\= \bigl\langle Y[T_g^{-1}\phi] \bigr\rangle_0\ .
\end{equation}
Here, we indicated the value of the coupling as a subscript on the correlator and also on
the symbol of the map~$T_g:\phi\mapsto \phi'[g,\phi]=T_g\phi$. 
The second equality expresses the distributivity of the map, i.e.~$T(\phi_1\phi_2)=(T\phi_1) \, (T\phi_2)$.
In the expectation values of~\eqref{Tdef} it is understood that all anticommuting degrees of freedom~$\psi$
have been integrated out.\footnote{
Auxiliary fields~$F$ may be kept as part of~$\phi$ or eliminated, see below.}
This means that the Nicolai map operates in a nonlocal bosonic theory with an action
\begin{equation}
S_g[\phi] \= S^{\mathrm{b}}_g[\phi] + \hbar\,S^{\mathrm{f}}_g[\phi]\ ,
\end{equation}
where the local part $S^{\mathrm{b}}_g$ is the bosonic piece of the original supersymmetric action
$S_{\textrm{\tiny SUSY}}[\phi,\psi]$ and $\exp\{\im S^{\mathrm{f}}_g\}$ arises from 
the path integral over the anticommuting fields in the partition function, both at coupling~$g$. 
In particular, $S^{\mathrm{f}}_g$ is in general nonlocal and suppressed by~$\hbar$. 
So far in the literature only theories with quadratic fermions have been considered, 
for which the integration over fermions can be formally carried out. As a result,
changing path-integral variables $Y\mapsto T_gY$ on the right-hand side of~\eqref{Tdef} and 
sorting powers of~$\hbar$, one recovers the original defining properties of the Nicolai map,
\begin{equation} \label{matchingold}
S^{\mathrm{b}}_0[T_g\phi] \= S^{\mathrm{b}}_g[\phi] \quad\und\quad
S^{\mathrm{f}}_0 -\im\,\tr\ln\sfrac{\delta T_g\phi}{\delta\phi} \= S^{\mathrm{f}}_g[\phi] \ ,
\end{equation}
which are called the `free-action' and `determinant-matching' property, respectively.
We note that $S^{\mathrm{f}}_0=S^{\mathrm{f}}_0[T_g\phi]$ is a constant 
since this functional does not depend on the bosonic fields.
The name `determinant-matching' was indeed coined because in theories with a flat target space 
integrating  the quadratic fermions out produces a fermionic
(or Matthews--Salam--Seiler) determinant, which  has to be matched by the Jacobian determinant
$\det\sfrac{\delta T_g\phi}{\delta\phi}$.
Powers of $\hbar$ are fully explicit because the fermionic determinant sums all fermionic loops 
in the bosonic background which is a one-loop exact operation. 
As a consequence, in the language of the `Nicolai rules' to construct the map, 
only tree diagrams appear, and for this reason it is sometimes considered a classical construction.

In this paper we will show that the assumption of quadratic fermions is not necessary; 
Nicolai maps can be constructed for supersymmetric actions with higher-order fermion terms as well.
The price to pay is quantum corrections to the Nicolai map and a more general dependence on~$\hbar$,
which upsets the conditions~\eqref{matchingold}.
In other words, the Nicolai map is a formal power series not only in~$g$ but also in~$\hbar$,
\begin{equation} \label{quantmap}
T_g\phi \= T_g^{(0)}\!\phi \ + \sum_{r=1}^\infty \hbar^r T_g^{(r)}\!\phi\ ,
\end{equation}
where $r$ counts the number of fermion loops in the diagrammatic representation.
With higher-order fermionic terms in the action, the path integral over the anticommuting fields
will produce an effective nonlocal action $\sum_r \hbar^r S_g^{(r)}[\phi]$
extending the previous one-loop result~$\hbar\,S^{\mathrm{f}}_g[\phi]$ to all orders in~$\hbar$.
Revising the argument leading from~\eqref{Tdef} to~\eqref{matchingold} 
we then find the combined identity
\begin{equation} \label{matchingnew}
S^{\mathrm{b}}_0[T_g\phi]\ +\ \hbar\,S^{\mathrm{f}}_0
\ -\ \im\hbar\,\tr\ln\sfrac{\delta T_g\phi}{\delta\phi}
\= S^{\mathrm{b}}_g[\phi]\ +\ \sum_r \hbar^r S_g^{(r)}[\phi]\ .
\end{equation}
Inserting~\eqref{quantmap} into~\eqref{matchingnew} and comparing powers in~$\hbar$
one obtains an infinite hierarchy of `Nicolai-map conditions'. The leading two represent the
tree-level and one-loop contributions and read
\begin{equation} \label{matching01}
S^{\mathrm{b}}_0[T_g^{(0)}\!\phi] \= S^{\mathrm{b}}_g[\phi] \quad\und\quad
S^{\mathrm{b}}_0[T_g\phi]\big|_{O(\hbar)} + S^{\mathrm{f}}_0 
- \im\,\tr\ln\sfrac{\delta T_g^{(0)}\!\phi}{\delta\phi} \= S_g^{(1)}[\phi] \ .
\end{equation}
Clearly, the `free-action condition' is still valid for the classical part of the Nicolai map,
but the `determinant-matching condition' receives a free-action contribution from the
one-loop correction to the map, and there are further conditions, each one balancing
expressions of a fixed loop order.

For any off-shell supersymmetric theory, there exists a formalism and a universal formula
which provides a formal power series expansion (in~$g$) of the map and 
its inverse~\cite{FL,DL1,L1,L2,ALMNPP,LR1,LR3,Lprague}.\footnote{
Under certain conditions the construction works also in the absence of off-shell supersymmetry,
e.g.~for super Yang--Mills theory in dimensions 6 and~10 in the Landau gauge~\cite{ALMNPP,MN,LR2,LR4}.}
The key player is the `coupling flow operator' 
\begin{equation}
R_g[\phi] \= \int\!\diff x\ \bigl(\pa_g T_g^{-1} \circ T_g \bigr) \phi(x)\,\frac{\delta}{\delta\phi(x)}\ ,
\end{equation}
where `$x$' stands for all coordinates the fields depend on.
This functional differential operator governs the infinitesimal Nicolai map,
\begin{equation} \label{localflow}
\pa_g \bigl\< Y[\phi] \bigr\>_g \= \bigl\< \bigl( \pa_g + R_g[\phi] \bigr) Y[\phi] \bigr\>_g\ .
\end{equation}
The first step towards its construction is the observation that, in off-shell supersymmetric chiral theories,
\begin{equation} \label{defDelta}
S_{\textrm{\tiny SUSY}}[\phi,\psi] \= \delta_\alpha \Delta_\alpha[\phi,\psi] 
\qquad\Rightarrow\qquad
\pa_g S_{\textrm{\tiny SUSY}}[\phi,\psi] \= \delta_\alpha \pa_g\Delta_\alpha[\phi,\psi] \ ,
\end{equation}
where $\alpha$ is a spinor index (we are being schematic here on the notation of spinors)
and $\Delta_\alpha$ is a certain anticommuting local functional.
Super Yang--Mills theories are more complicated because $[\pa_g,\delta_\alpha]\neq0$, 
and we exclude them in the following.
Employing~(\ref{defDelta}) and the supersymmetric Ward identity we integrate out the anticommuting variables 
to read off the coupling flow operator 
\begin{equation} \label{flowop}
R_g[\phi] \= \frac{\im}{\hbar}\,\pa_g\bcontraction{}{\Delta}{_\alpha[\phi]\ }{\delta} \Delta_\alpha[\phi]\ \delta_\alpha \=
\frac{\im}{\hbar}\int\!\diff x\ \pa_g\bcontraction{}{\Delta}{_\alpha[\phi]\ }{\delta} \Delta_\alpha[\phi]\ \delta_\alpha \phi(x)\ 
\frac{\delta}{\delta\phi(x)}\ ,
\end{equation}
where the contraction indicates a fermionic expectation value to be taken. 

Exponentiating the action of $\pa_g{+}R_g$ generates the (finite-flow) inverse Nicolai map.
Alternatively, a $g$-ordered exponential of ${-}\!\!\int_0^g\diff{g'}\,R_{g'}$ directly yields the Nicolai map.
In any case, the $R_g$ action has to be iterated, $R_{g_1}R_{g_2}\cdots R_{g_s}\phi$,
which grafts full fermionic $2k$-point functions onto previously generated diagrams.
For the Wess--Zumino model (quadratic in the fermions) this produces fermionic tree diagrams, dressed with bosonic `leaves'.
Still, fermion loops remain absent.
For nonlinear sigma models, however, the graphical expansion of the Nicolai map will feature fermionic
trees with all kinds of fermion loops embedded. Thus it can no longer be considered a classical map.
Nevertheless, at any power of the coupling~$g$ a finite number of diagrams contributes to the Nicolai map,
and we may still employ the universal formula to write it down.
In the following, this will be demonstrated for four-dimensional supersymmetric sigma models, first in general
and second for supersymmetric $\C{\rm P}^N$ models. The latter, being maximally symmetric,  
admit  a formulation with only quadratic fermions via the introduction of an auxiliary vector field. 
We do not fully explore its implications here, as this falls outside the scope of the present paper, 
which presents a more general construction.

Furthermore, for the scope of this paper, relating the perturbative Nicolai map to the standard perturbative (Feynman) expansion 
solely concerns the generation of diagrams. These are in general divergent and in most applications require regularization. 
We assume that  it has been been done in an appropriate way, e.g.\ via dimensional regularization or with a UV cutoff.\footnote{
In particular, the perturbative non-renormalizability of the nonlinear sigma models described here is not of concern. }
These aspects, and more generally the interplay between the Nicolai map and regularization or renormalization, 
deserve further study. 

There are several  ways in which the work presented here can be further expanded or generalized.  
The implications of the auxiliary vector in the $\C {\rm P}^N$ model deserve further study.\footnote{
For example, it might be possible to improve the construction presented here by choosing a different normalization for the fields, 
which is crucial in the construction of the map in the super Yang--Mills case \cite{LR1}. Work in this direction is ongoing. }
It would  be interesting to work out explicitly additional orders of the Nicolai map for \(\C{\rm P}^N \) models 
presented here and study in detail the higher-loop identities ~\eqref{matchingnew}. 
Conceivably, then, the formalism presented here can be naturally extended to gauge theories. 
Finally, a more ambitious goal is  the application of the Nicolai map to supergravity, 
which has been excluded so far owing to the four-fermion contributions, 
with the potential application of shedding light on  its UV behaviour.

The rest of the paper is organized as follows.
Section~\ref{s::nlsm} collects some general expressions for supersymmetric nonlinear sigma models 
and its coupling flow operator. 
In Section~\ref{s::CPN} we specialize to \(\C \mathrm P^N\) and construct the associated Nicolai map in perturbation theory. 
A possibility of eliminating the four-fermion interactions via auxiliary (Hubbard--Stratonovich) scalars or vectors
is discussed in Section~\ref{s::A}. We then comment on the superfield origin of an auxiliary vector in \(\C \mathrm P^N\) models
and its relevance for the construction of the perturbative Nicolai map. 

\section{Supersymmetric nonlinear sigma model}\label{s::nlsm}

\noindent
The prototypical supersymmetric field theory with higher-than-quadratic fermion terms in the action is
the supersymmetric nonlinear sigma model in $(3{+}1)$-dimensional Minkowski space \cite{bertolini}, 
which is characterized by a K\"ahler potential ${\rm K}(\Phi_a^\dagger,\Phi^a)$
and a superpotential ${\rm W}(\Phi^a)$ for a collection 
\begin{equation}
\{\Phi^a\} \= \{\Phi^0,\Phi^A\} \= \{\Phi^0,\Phi^1,\Phi^2,\ldots,\Phi^N\}
\end{equation}
of $N{+}1$ chiral superfields, \ie $a,b,c,\ldots=0,1,\ldots,N$ and $A,B,C,\ldots=1,\ldots,N$.
Complex conjugation raises or lowers target-space indices.
Adopting the Wess--Bagger notation~\cite{wessbagger} their component expansion (in $x$ coordinates) reads
\begin{equation}
\Phi^a\= \phi^a
+ \im \theta \sigma^m \bar\theta\, \partial_m \phi^a
+ \tfrac14 \theta^2 \bar \theta^2\,\Box \phi^a
+ \sqrt 2 \theta\, \psi^a
- \tfrac{\im}{\sqrt 2}  \theta^2 \partial_m \psi^a \sigma^m \bar\theta
+ \theta^2 F^a
\ =:\ \phi^a + \Xi ^a\ ,
\end{equation}
where Weyl spinor indices $\alpha,\dot\alpha=1,2$ are suppressed.
We note that $\Xi^3=0$.
For the purpose of this paper the superpotential is not essential (although it can easily be added),
and thus we omit any $F$-term in the action and restrict ourselves to the K\"ahler potential
in the supersymmetric D-term action
\begin{equation} \label{Kaction}
S_{\textrm{\tiny SUSY}} \ = \int\!\diff^4x \ \Lagr \qquad\textrm{with}\qquad
\Lagr \= \int\!\diff^2\theta\,\diff^2\bar\theta \ {\rm K}(\Phi^\dagger_a ,\Phi^a)\ .
\end{equation}
We further define the functional $\Delta$ given by
\begin{equation}
\Delta \ = \int\!\diff^4x \ \Delt \qquad\textrm{with}\qquad
\Delt \= \int\!\diff\theta\,\diff^2\bar\theta \ {\rm K}(\Phi^\dagger_a ,\Phi^a)\ ,
\end{equation}
so that
\begin{equation}
\Lagr \= -\tfrac14 \bigl( 
\tfrac{1+\kappa}{2}\,\delta^\alpha \Delt_\alpha 
+ \tfrac{1-\kappa}{2}\,\bar\delta_{\dot\alpha} \bar\Delt^{\dot\alpha} \bigr)
\ +\ \textrm{total derivative}\ ,
\end{equation}
with a free parameter~$\kappa$ controlling the relative weight
of the chiral against antichiral contribution.\footnote{
This freedom is   R symmetry: a D-term can be reached in two ways,
by applying either $\delta^\alpha$ to $\Delt_\alpha$ or $\bar\delta_{\dot\alpha}$ to $\bar\Delt^{\dot\alpha}$.}

We introduce the notation
\begin{equation}
K := {\rm K}(\phi_a^*,\phi^a) \und
K_{ab\ldots}^{cd\ldots} 
:= \tfrac{\pa}{\pa\phi^a}\tfrac{\pa}{\pa\phi^b}\cdots\tfrac{\pa}{\pa\phi_c^*}\tfrac{\pa}{\pa\phi_d^*}\cdots K\ .
\end{equation}
Expanding
\begin{equation}
\begin{aligned}
{\rm K}
&\= K + K_a\,\Xi^a + K^a\,\bar \Xi_a 
+  K^a_b\,\Xi^b  \bar \Xi_a + \tfrac 12   K_{ab }\,\Xi^a\Xi^b + \tfrac 12   K^{ab }\,\bar\Xi_a\bar\Xi_b 
\\ & \quad
+ \tfrac12  K^a_{bc}\,\Xi^b\Xi^c  \bar \Xi_a
+  \tfrac 12   K^{bc }_a\,\Xi^a \bar\Xi_b \bar\Xi_c 
+ \tfrac 14 K^{ab}_{cd}\,\Xi^a \Xi^b \bar\Xi_c \bar\Xi_d 
\\[4pt]
&\= \ldots  + \bar\theta^2 \theta\, \Delt +   \theta^2 \bar \theta\, \bar \Delt
+ \theta^2 \bar \theta^2 \, \Lagr
\end{aligned}
\end{equation}
and employing the identity
\begin{equation}
\Box\,K \= 
K^{ab} \, \partial_m \phi_a^*\partial^m \phi_b^*
+ K_{ab} \, \partial_m \phi^a\partial^m \phi^b
+ 2 K^a_b \, \partial_m \phi_a^*\partial^m \phi^b
+ K^a \, \Box \phi^*_a
+ K_a \, \Box \phi^a \ ,
\end{equation}

we read off the component lagrangian
\begin{equation} \label{lagr}
\begin{aligned}
\Lagr &\= K^b_a
\left[ 
- \partial^m \phi^a \partial_m \phi^*_b 
- \tfrac{\im}{2} \psi^a \sigma^m \partial_m \bar \psi_ b
+ \tfrac{\im}{2} \partial_m   \psi  ^a \sigma^m \bar \psi_b
+ F^a F^*_b
\right]
\\ & \quad {}
+ \tfrac12 K_{ab}^c
\left[ 
\im \psi^b \sigma^m \bar \psi_c\,\partial_m \phi^a 
- \psi^a \psi^b F^*_c
\right]
+\tfrac12  K^{ab}_c
\left[ 
\im \bar \psi_b\, \bar \sigma^m   \psi^c \partial_m \phi_a 
- \bar \psi_a \bar \psi_b \, F^c
\right]
\\
& \quad {}
+ \tfrac 14 K_{cd}^{ab}  \  \bar \psi_a \bar \psi_b  \ \psi^c \psi^d 
\\[4pt]
&\= K^b_a
\left[ 
- \partial^m \phi^a \partial_m \phi^*_b 
- \tfrac{\im}{2} \psi^a \sigma^m \nabla_m \bar \psi_ b
+ \tfrac{\im}{2} \nabla_m   \psi  ^a \sigma^m \bar \psi_b
+ F^a F^*_b
\right]
\\
& \quad {}
- \tfrac12 K_{ab}^c\ \psi^a \psi^b F^*_c 
- \tfrac12 K^{ab}_c\ \bar \psi_a \bar \psi_b \, F^c  
+ \tfrac 14 K_{cd}^{ab}  \  \bar \psi_a \bar \psi_b  \ \psi^c \psi^d
\end{aligned}
\end{equation}
with  target-space covariant derivative on spinors defined as
\begin{equation}
K^b_a\ \nabla_m \psi^a = K^b_a\,\partial_m \psi^a + K^b_{cd}\,\partial_m \phi^c \psi^d\ ,
\end{equation}
and we also obtain the penultimate component
\begin{equation} \label{delt}
\Delt \=  
\sqrt{2}\,K_a^b \big[ \psi^a F^*_b - \im \sigma^m \bar \psi_b \, \partial_m\phi^a \big] 
+ \tfrac{1}{\sqrt 2} K^{bc }_a\;\psi^a \;\bar \psi_b \bar \psi_c\ ,
\end{equation}
which are both manifestly invariant under K\"ahler transformations
$\rm K \mapsto K + \Lambda + \Lambda^\dagger$.

In order to simplify the coupling-flow operator, it is advisable to also integrate out
the auxiliary~$F^a$ by inserting its classical value
\begin{equation} \label{Feom}
K^b_a\,F^*_b \= \tfrac12 K_a^{cd}\ \bar\psi_c \bar\psi_d 
\qquad\Rightarrow\qquad
F^a \= \tfrac12 (K^{-1})^a_b \, K^b_{cd}\ \psi^c\psi^d
\end{equation}
back into $\Lagr$ and $\Delt$, which produces
\begin{equation}
\Lagr \= K^b_a
\left[ 
- \partial^m \phi^a \partial_m \phi^*_b 
- \tfrac{\im}{2} \psi^a \sigma^m \nabla_m \bar \psi_ b
+ \tfrac{\im}{2} \nabla_m   \psi  ^a \sigma_m \bar \psi_b
\right]
\ +\ \tfrac14 R^{ab}_{cd} \  \bar \psi_a \bar \psi_b  \ \psi^c \psi^d
\end{equation}
with the Riemann tensor
\begin{equation} \label{riemann}
R^{ab}_{cd} \= K_{cd}^{ab}   -  K^r_{cd}  (K^{-1})_r^s K_s^{ab}
\end{equation}
as well as 
\begin{equation} \label{delton}
\Delt \=  -\im \sqrt{2}\,K_a^b\ \sigma^m \bar \psi_b\, \partial_m\phi^a
\ +\ \sqrt 2 \, K^{bc }_a\;\psi^a \;\bar \psi_b \bar \psi_c \ .
\end{equation}
We note that the invertibility of the K\"ahler metric is not needed
in the final expression for the coupling-flow operator. 
Still, as we shall see, if redundant target coordinates are employed
we must constrain them (`fix a gauge') and also introduce a coupling~$g$ 
into the K\"ahler potential~$K$ in order to obtain 
a perturbative expansion for the Nicolai map.

Suppose now that $S_{\textrm{\tiny SUSY}}$ features terms of order $2k$ in the $\psi$ fields, with $k=1,2,\ldots$.
Then, $\pa_g\Delt_\alpha$ has terms of the form $\psi^{2k-1}$.
Since $\delta_\alpha$ acting on $\phi$ is linear in~$\psi$, the contraction in
\begin{equation}  \label{Rg}
R_g \= -\tfrac{\im}{4\hbar} \int\!\diff{y}\!\int\!\diff{x}\ \Bigl\{
\tfrac{1+\kappa}{2}\,\pa_g\bcontraction{}{\Delt}{_\alpha(y)\ }{\delta}
\Delt_\alpha(y)\ \delta^{\alpha}(x)\ +\
\tfrac{1-\kappa}{2}\,\pa_g\bcontraction{}{\bar\Delt}{^{\dot\alpha}(y)\ }{\bar\delta}
\bar\Delt^{\dot\alpha}(y)\ \bar\delta_{\dot\alpha}(x) \Bigr\}
\end{equation}
signifies a fermionic $2k$-point function in the $\phi$~background,
with $2k{-}1$ legs fused.\footnote{
$\pa_g{\cal M}$ must be computed from~\eqref{delt},
before eliminating auxiliaries, because \eqref{Rg} relies on off-shell supersymmetry.}
For a flat target space, we look at Wess--Zumino chiral models, and fermions appear only quadratically in the action.
Thus, $k{=}1$ implies only a full fermion propagator in~\eqref{flowop},
\begin{equation} \label{2point}
\bcontraction{}{\bar\psi}{_b^{\dot\beta}(x')\ }{\psi} 
\bar\psi_b^{\dot\beta}(x')\ \psi^a_\alpha(x) \ \equiv\
\hbar\,(G_2^{(g)})^{a\dot\beta}_{b\alpha}(x'\!,x)\ ,
\end{equation}
which contains all Feynman diagrams connected by a single fermion line with external $\phi$~legs.
Expanding the full fermion propagator in powers of~$g$, one obtains chains of free fermion propagators,
with vertices encoding the coupling to the bosonic background. No fermion loops arise.
In case of a curved target space, instead, we face a nonlinear sigma model, whence $S_{\textrm{\tiny SUSY}}$
has terms quadratic ($k{=}1$) as well as quartic ($k{=}2$) in the fermions. In this case,
a $\psi^3$ contribution in $\Delt_\alpha$ produces a correlator of a composite~$\psi^3$ with another~$\psi$ 
in the bosonic background, which then occurs in~\eqref{flowop},
\begin{equation} \label{4point}
\bcontraction{}{\psi}{^b_\beta(x')}{\!\psi} 
\bcontraction{\psi^b_\beta(x')}{\bar\psi}{_c^{\dot\gamma}(x')}{\!\psi}
\bcontraction{\psi^b_\beta(x')\bar\psi^c_{\dot\gamma}(x')}{\bar\psi}{_d^{\dot\delta}(x')\ }{\!\psi} 
\psi^b_\beta(x')\bar\psi_c^{\dot\gamma}(x')\bar\psi_d^{\dot\delta}(x')\ \psi^a_\alpha(x) \ \equiv\
\hbar^2 (G_4^{(g)})^{ba\dot\gamma\dot\delta}_{cd\beta\alpha}(x'\!,x)\ ,
\end{equation}
which contains all Feynman diagrams connecting a triple fermion vertex with another fermion. 
We note that both $G_2^{(g)}$ and $G_4^{(g)}$ include diagrams with fermion loops
generated by the four-fermion coupling.
Expanding in powers of~$g$ and taking into account the $\psi^4$ interaction in $S_{\textrm{\tiny SUSY}}$,
we encounter fermion loop diagrams in the graphical expansion 
of the coupling flow operator and hence of the Nicolai map. 

\section{Supersymmetric $\C{\rm P}^N$ model}\label{s::CPN}

\noindent
Let us become more concrete and specialize to a maximally symmetric K\"ahler target,
namely the complex projective space
$\C{\rm P}^N\simeq\frac{\textrm{SU}(N{+}1)}{\textrm{SU}(N)\times\textrm{U}(1)}$. 
It is embedded into $\C^{N+1}\ni\phi^a$
by the identification $\phi^a\sim\lambda\,\phi^a$ with $\lambda\in\C^*$, 
which yields the K\"ahler potential
\begin{equation} \label{Kcpn}
K(\phi^*_a,\phi^a) \= \frac{\mu^2}{g}\,\log\Bigl[\frac{g}{\mu^2}\,\phi^*\phi \Bigr]
\qquad\textrm{with}\qquad \phi^*\phi \equiv \phi^*_a \phi^a\ ,
\end{equation}
where \(g\) is a dimensionless coupling,
and $\mu$ is a mass parameter accounting for the dimensionality of~$\phi$.\footnote{
There really is only one (dimensionful) parameter $M^2=\mu^2/g$; 
we introduce the dimensionless coupling~$g$ only for later convenience.
Also, $\mu^2$ need not be positive.}
The superfield extension identifies $\Phi^a\sim\Lambda\,\Phi^a$ with 
a complex chiral superfield~$\Lambda$.
For later use, we introduce the abbreviations
\begin{equation}
f_g^{-1} \ :=\ \tfrac{g}{\mu^2}\,\phi^*\phi 
\qquad\und\qquad
\Pi_a^b \ :=\ \delta_a^b - \tfrac{\phi^b \phi^*_a}{\phi^*\phi}\ ,
\end{equation}
in terms of which the derivatives of~$K$ are as follows,
\begin{equation} \label{Kders}
K_a^b = f_g\,\Pi_a^b \, ,\qquad\quad
K_a^{bc} = -2 f_g^2 \tfrac{g}{\mu^2}\,\phi^{(b}\Pi^{c)}_a \,, \quad\qquad
K_{ab}^{cd} = 4 f_g^3 \tfrac{g^2}{\mu^4}\,\phi^{(c}\Pi^{d)}_{(a}\phi^*_{b)}
- 2 f_g^2 \tfrac{g}{\mu^2}\,\Pi^{(c}_{(a}\Pi^{d)}_{b)}\ .
\end{equation}
We remark that $\Pi^b_a\,\phi^a=0=\phi^*_b\,\Pi^b_a$ implements the projection
transversal to the `radial direction' of the identification $\phi^a\sim\lambda\,\phi^a$.

As it stands, this K\"ahler potential is singular in the $g\to0$ limit.
In order to set up a perturbation theory around flat~$\C^N$, we need to select a point
on~$\C{\rm P}^N$ and pick coordinates $\phi^A\in\C^N$ centered around it.
An appropriate superfield choice fixing the above redundancy is
\begin{equation} \label{gauge}
\Phi^0 = \sqrt{\tfrac{\mu^2}{g}}  \qquad\Leftrightarrow\qquad
\bigl\{ \phi^0 , \psi^0, F^0 \bigr\} \= \bigl\{ \sqrt{\tfrac{\mu^2}{g}} , 0, 0 \bigr\}\ ,
\end{equation}
which leaves us with $N$~proper (super)coordinates
\begin{equation}
\Phi^A \= \phi^A
+ \im \theta \sigma^m \bar\theta\, \partial_m \phi^A
+ \tfrac14 \theta^2 \bar \theta^2\,\Box \phi^A
+ \sqrt 2 \theta\, \psi^A
- \tfrac{\im}{\sqrt 2}  \theta^2 \partial_m \psi^A \sigma^m \bar\theta
+ \theta^2 F^A
\end{equation}
and a well-behaved K\"ahler potential
\begin{equation}
K(\phi^*_A,\phi^A) \= \frac{\mu^2}{g}\,\log\Bigl[ 1 + \frac{g}{\mu^2}\,\hat\phi^*\hat\phi \Bigr]
\= \hat\phi^*\hat\phi\ -\ \tfrac12 \tfrac{g}{\mu^2}\bigl(\hat\phi^*\hat\phi\bigr)^2 +\ O(g^2)
\qquad\textrm{with}\qquad \hat\phi^*\hat\phi \equiv \phi^*_A \phi^A \ .
\end{equation}
Observing that \eqref{gauge} effectively reduces the summation ranges in \eqref{lagr} and \eqref{delt} 
to $1,\ldots,N$, \ie $a,b,\ldots\to A,B,\ldots$, we only need to insert
\begin{equation} \label{Khatders}
K_A^B = f_g\,\Pi_A^B \ ,\quad
K_A^{BC} = -2 f_g^2 \tfrac{g}{\mu^2}\,\phi^{(B}\Pi^{C)}_A \ ,\quad
K_{AB}^{CD} = 4 f_g^3 \tfrac{g^2}{\mu^4}\,\phi^{(C}\Pi^{D)}_{(A}\phi^*_{B)} 
- 2 f_g^2 \tfrac{g}{\mu^2}\,\Pi^{(C}_{(A}\Pi^{D)}_{B)}
\end{equation}
with \ $f_g^{-1}=1+\tfrac{g}{\mu^2}\,\hat\phi^*\hat\phi$ \ and
\begin{equation}
\Pi^B_A \= \delta^B_A - f_g\,\tfrac{g}{\mu^2}\,\phi^B \phi^*_A
\qquad\Rightarrow\qquad
\Pi^B_A\,\phi^A = f_g\,\phi^B \und
\phi^*_B\,\Pi^B_A = f_g\,\phi^*_A \ .
\end{equation}

The elimination of the auxiliary~$F$ via \eqref{Feom} commutes with the coordinate choice for~$\Delt$,
\begin{equation} \label{deltoffon}
\begin{aligned}
\Delt &\= 
-\im \sqrt{2}\,K_A^B\ \slashed{\pa}\phi^A\, \bar\psi_B
\ +\ \sqrt{2}\,K_A^B\ \psi^A\,F^*_B
\ +\ \tfrac{1}{\sqrt 2} \, K^{BC }_A\;\psi^A \;\bar\psi_B \bar\psi_C 
\\[4pt]
&\=  -\im \sqrt{2}\,K_A^B\ \slashed{\pa}\phi^A\, \bar\psi_B
\ +\ \sqrt 2 \, K^{BC }_A\;\psi^A \;\bar\psi_B \bar\psi_C \\[4pt]
&\= -\im \sqrt{2}\,f_g\,\Pi^B_A \,\slashed{\pa}\phi^A\, \bar\psi_B
\ -\  2 \sqrt 2\,f_g^2\tfrac{g}{\mu^2}\, \phi^C\, \Pi^B_A\,\psi^A \ \bar\psi_B \bar \psi_C\ ,
\end{aligned}
\end{equation}
but not so for the lagrangian \eqref{lagr} because \eqref{riemann} requires
invertibility of the K\"ahler metric~$K^b_a$, which is degenerate in the redundant coordinates.
However, after choosing \eqref{gauge} we simply obtain
\begin{equation}
K^B_A\,F^*_B \= \tfrac12 K_A^{CD}\ \bar\psi_C \bar\psi_D 
\qquad\Rightarrow\qquad
F^A \= \tfrac12 ({K}^{-1})^A_B \, K^B_{CD}\ \psi^C\psi^D
\quad\textrm{with}\quad
({K}^{-1})^A_B\,K^B_C=\delta^A_C
\end{equation}
with an inverse ${K}^{-1}$ on~$\C{\rm P}^N$
and hence a four-fermion interaction
\begin{equation} \label{4fermi}
\tfrac14 {R}^{AB}_{CD}  \  \bar \psi_A \bar \psi_B  \ \psi^C \psi^D
\qquad\textrm{with}\qquad
{R}^{AB}_{CD} \= K_{CD}^{AB}   -  K^R_{CD}  ({K}^{-1})_R^S K_S^{AB}\ .
\end{equation}
The $\C{\rm P}^N$ Riemannn tensor computes to (c.f.~\eqref{Khatders})
\begin{equation} \label{Rfact}
{R}^{AB}_{CD} \= - 2 f_g^2 \tfrac{g}{\mu^2}\,\Pi^{(A}_{(C}\Pi^{B)}_{D)}
\= -\tfrac{g}{\mu^2} \bigl( K^A_C K^B_D + K^A_D K^B_C \bigr)\ .
\end{equation}
Working out the details one arrives at
\begin{equation} \label{LagrCPN}
\begin{aligned}
\Lagr &\= 
f_g\,\Pi^{ B}_A \bigl[
- \partial_m \phi^A\, \partial^m \phi^*_B   
- \tfrac{\im}{2} (\psi^A \slashed{\pa}  \bar \psi_B
+\bar\psi_B \bar{\slashed{\pa}} \psi^A)
+\tfrac{\im}{2}\, f_g \tfrac{g}{\mu^2}\,
( \psi^A \phi^C \slashed{\pa} \phi^*_B {\bar \psi}_C
+ \bar\psi_B \phi^*_C \bar{\slashed{\pa}}\phi^A \psi^C )
\bigr]
\\[4pt] & \quad
+\tfrac{\im}{2}\, f_g^2 \tfrac{g}{\mu^2} \,\Pi^B_A\,
\psi^A (\phi^C\slashed{\pa}\phi^*_C-\phi^*_C\slashed{\pa}\phi^C)\,\bar\psi_B
+ \tfrac 14 \,f_g^2 \tfrac{g}{\mu^2}
\bigl[\psi^C \sigma_m \Pi^A_C \bar \psi_A  \bigr]
\bigl[\psi^D \sigma^m \Pi^B_D \bar \psi_B \bigr] \ ,
\end{aligned}
\end{equation}
where we have employed the Fierz identity
\begin{equation} \label{fierz}
[ \bar\psi_A \bar\psi_B ]  \,[ \psi^C\psi^D ] \= - \tfrac 12 
[ \psi^C \sigma_m \bar\psi_A ] \, [ \psi^D \sigma^m \bar\psi_B ] \ .
\end{equation}

We now have all the ingredients for constructing the Nicolai map.
Remembering that
\begin{equation}
\delta_\alpha \phi^A = \sqrt{2}\,\psi^A_\alpha \quad ,\qquad
\delta_\alpha \phi^*_A = 0 \quad ,\qquad
{\bar\delta}^{\dot\alpha} \phi^A = 0 \quad ,\qquad
{\bar\delta}^{\dot\alpha} \phi^*_A = \sqrt{2}\,{\bar\psi}_A^{\dot\alpha}
\end{equation}
the coupling-flow operator takes the form~\footnote{
To be precise, in $\pa_g{\cal M}_\alpha$ the auxiliary~$F$ has been eliminated only after
applying the $g$-derivative to the first line of~\eqref{deltoffon}.}
\begin{equation}
\begin{aligned}
R_g &\= -\tfrac{\im}{4\hbar}\sqrt{2}\int\!\diff{y}\!\int\!\diff{x}\ \Bigl\{
\tfrac{1+\kappa}{2}\,\pa_g\bcontraction{}{\Delt}{_\alpha(y)\ }{\psi}
\Delt_\alpha(y)\ \psi^{A\alpha}(x)\,\tfrac{\delta}{\delta\phi^A(x)}\ +\
\tfrac{1-\kappa}{2}\,\pa_g\bcontraction{}{\bar\Delt}{^{\dot\alpha}(y)\ }{\bar\psi}
\bar\Delt^{\dot\alpha}(y)\ \bar\psi_{A\dot\alpha}(x)\,\tfrac{\delta}{\delta\phi^*_A(x)}
\Bigr\} \\[4pt]
&\= \tfrac{1}{2\mu^2\hbar} \int\!\diff{y}\!\int\!\diff{x}\ \tfrac{1+\kappa}{2}\,f_g^2\Bigl\{ 
\bigl(\hat\phi^*\hat\phi\,\delta_A^B+(2f_g{-}1)\,\phi^B\phi_A^*\bigr)\,
\slashed{\pa}_{\alpha\dot\alpha} \phi^A(y)\,
\bcontraction{}{\bar\psi}{_B^{\dot\alpha}(y)\  }{\psi}
\bar\psi_B^{\dot\alpha}(y)\ \psi^{D\alpha}(x)\,\tfrac{\delta}{\delta\phi^D(x)} \\
&\qquad\qquad 
+\im\bigl((3f_g{-}2)\delta_A^B-\tfrac{g}{\mu^2}f_g(5f_g{-}2)\,\phi^B\phi_A^*\bigr)\,\phi^C(y)\,
\bcontraction{}{\psi}{^A_\alpha(y)\,}{\!\bar\psi}
\bcontraction{\psi^A_\alpha(y)\,}{\bar\psi}{_{B\dot\alpha}(y)\,}{\!\bar\psi}
\bcontraction{\psi^A_\alpha(y)\,\bar\psi_{B\dot\alpha}(y)\,}{\bar\psi}{_C^{\dot\alpha}(y)\ }{\!\psi}
\psi^A_\alpha(y)\,\bar\psi_{B\dot\alpha}(y)\,\bar\psi_C^{\dot\alpha}(y)\
\psi^{D\alpha}(x)\,\tfrac{\delta}{\delta\phi^D(x)} \Bigr\} \\
&\ \ \ +\tfrac{1}{2\mu^2\hbar} \int\!\diff{y}\!\int\!\diff{x}\ \tfrac{1-\kappa}{2}\,f_g^2\Bigl\{ 
\bigl(\hat\phi^*\hat\phi\,\delta_A^B+(2f_g{-}1)\,\phi^B\phi_A^*\bigr)\,
\bar{\slashed{\pa}}^{\dot\alpha\alpha} \phi^*_B(y)\,
\bcontraction{}{\psi}{^A_\alpha(y)\ }{\bar\psi}
\psi^A_\alpha(y)\ \bar\psi_{D\dot\alpha}(x)\,\tfrac{\delta}{\delta\phi^*_D(x)} \\
&\qquad\qquad 
+\im\bigl((3f_g{-}2)\delta_A^B-\tfrac{g}{\mu^2}f_g(5f_g{-}2)\,\phi^B\phi_A^*\bigr)\,\phi^*_C(y)\,
\bcontraction{}{\bar\psi}{_B^{\dot\alpha}(y)\,}{\!\psi}
\bcontraction{\bar\psi_B^{\dot\alpha}(y)\,}{\psi}{^{A\alpha}(y)\,}{\psi}
\bcontraction{\bar\psi_B^{\dot\alpha}(y)\,\psi^{A\alpha}(y)\,}{\psi}{^C_\alpha(y)\ }{\!\bar\psi}
\bar\psi_B^{\dot\alpha}(y)\,\psi^{A\alpha}(y)\,\psi^C_\alpha(y)\
\bar\psi_{D\dot\alpha}(x)\,\tfrac{\delta}{\delta\phi^*_D(x)} \Bigr\} 
\end{aligned}
\end{equation}
with full fermionic correlators indicated by the contractions, see \eqref{2point} and~\eqref{4point}.

Let us take a look at first order in the coupling~$g$,
\begin{equation}
(T_g \phi)^A(x) \= \phi^A(x) \ -\ g\,(R_g\phi)|_{g=0}^A(x)\ +\ O(g^2)\ .
\end{equation}
We compute
\begin{equation}
\begin{aligned}
(R_g\phi)|_{g=0}^A(x) 
&\= \tfrac{1+\kappa}{4\hbar\mu^2} \!\int\!\diff{y}\ \Bigl\{ 
\phi^*_B
(\phi^{B}\slashed{\pa}_{\alpha\dot\alpha}\phi^{C}
+\phi^{C}\slashed{\pa}_{\alpha\dot\alpha}\phi^{B})\,
\bcontraction{}{\bar\psi}{_C^{\dot\alpha}(y)\  }{\psi}
\bar\psi_C^{\dot\alpha}(y)\ \psi^{A\alpha}(x) 
+ \im \,\phi^C\,
\bcontraction{}{\psi}{^B_\alpha\,}{\!\bar\psi}
\bcontraction{\psi^B_\alpha\,}{\bar\psi}{_{B\dot\alpha}\,}{\!\bar\psi}
\bcontraction{\psi^B_\alpha\,\bar\psi_{B\dot\alpha}\,}{\bar\psi}{_C^{\dot\alpha}(y)\ }{\!\psi}
\psi^B_\alpha\,\bar\psi_{B\dot\alpha}\,\bar\psi_C^{\dot\alpha}(y)\
\psi^{A\alpha}(x) \Bigr\} \ , \\
(R_g\phi^*)|^{g=0}_A(x) 
&\= \tfrac{1-\kappa}{4\hbar\mu^2} \!\int\!\diff{y}\ \Bigl\{ 
\phi^B
(\phi^*_{B}\,\bar{\slashed{\pa}}^{\dot\alpha\alpha}\!\phi^*_{C}
+\phi^*_{C}\,\bar{\slashed{\pa}}^{\dot\alpha\alpha}\!\phi^*_{B})\,
\bcontraction{}{\psi}{^C_\alpha(y)\ }{\bar\psi}
\psi^C_\alpha(y)\ \bar\psi_{A\dot\alpha}(x) 
+ \im \,\phi^*_C\,
\bcontraction{}{\bar\psi}{_B^{\dot\alpha}\,}{\!\psi}
\bcontraction{\bar\psi_B^{\dot\alpha}\,}{\psi}{^{B\alpha}\,}{\!\psi}
\bcontraction{\bar\psi_B^{\dot\alpha}\,\psi^{B\alpha}\,}{\psi}{^C_\alpha(y)\ }{\!\bar\psi}
\bar\psi_B^{\dot\alpha}\,\psi^{B\alpha}\,\psi^C_\alpha(y)\
\bar\psi_{A\dot\alpha}(x) \Bigr\} \ ,
\end{aligned}
\end{equation}
where now the contractions are free-field ones,
\begin{equation}
\begin{aligned}
\sigma^m_{\alpha\dot\alpha}\
\bcontraction{}{\bar\psi}{_C^{\dot\alpha}(y)\  }{\psi}
\bar\psi_C^{\dot\alpha}(y)\ \psi^{A\alpha}(x) &\=
-2\,\hbar\,\delta_C^A\,\pa^m \Box^{-1}(y{-}x) \=
\bar\sigma^{m\dot\alpha\alpha}\
\bcontraction{}{\psi}{^A_\alpha(y)\ }{\bar\psi}
\psi^A_\alpha(y)\ \bar\psi_{B\dot\alpha}(x)
\ ,\\[4pt]
\bcontraction{}{\psi}{^B_{\alpha}\,}{\bar\psi}
\bcontraction{\psi^B_{\alpha}\,}{\bar\psi}{_{B\dot\alpha}\,}{\!\bar\psi}
\bcontraction{\psi^B_{\alpha}\,\bar\psi_{B\dot\alpha}\,}{\bar\psi}{_C^{\dot\alpha}(y)\ }{\!\psi}
\psi^B_{\alpha}\,\bar\psi_{B\dot\alpha}\,\bar\psi_C^{\dot\alpha}(y)\
\psi^{A\alpha}(x) &\=
2\,\hbar^2 (N{+}1)\,\delta_C^A\,\pa_m\Box^{-1}(y{-}z)\,\pa^m\Box^{-1}(y{-}x)\big|_{z=y}\ ,
\end{aligned}
\end{equation}
and arrive at  the classical map
\begin{equation}
\begin{aligned}
(T_g^{(0)}\!\phi)^A(x) &\= \phi^A(x)\ +\ \tfrac{1+\kappa}{2}\,\tfrac{g}{\mu^2} \int\!\diff{y}\ 
\phi^*_B\,\pa_m(\phi^{B}\phi^{A})(y)\,\pa^m \Box^{-1}(y{-}x)\ +\ O(g^2)\ , \\
(T_g^{(0)}\!\phi^*)_A(x) &\= \phi^*_A(x)\ +\ \tfrac{1-\kappa}{2}\,\tfrac{g}{\mu^2} \int\!\diff{y}\ 
\phi^B\,\pa_m(\phi^*_{B}\phi^*_{A})(y)\,\pa^m \Box^{-1}(y{-}x)\ +\ O(g^2)\ ,
\end{aligned}
\end{equation}
while the four-fermion contraction yields the leading (or one-loop) quantum contribution
\begin{equation}
\begin{aligned}
(T_g^{(1)}\!\phi)^A(x) &\= -\im\tfrac{g}{\mu^2}\tfrac{1{+}\kappa}{2}(N{+}1)\int\!\diff{y}\ 
\phi^A(y)\,\pa_m\Box^{-1}(y{-}z)\,\pa^m\Box^{-1}(y{-}x) \big|_{z=y}\ +\ O(g^2) \ , \\[4pt]
(T_g^{(1)}\!\phi^*)_A(x) &\= -\im\tfrac{g}{\mu^2}\tfrac{1{-}\kappa}{2}(N{+}1)\int\!\diff{y}\ 
\phi^*_A(y)\,\pa_m\Box^{-1}(y{-}z)\,\pa^m\Box^{-1}(y{-}x) \big|_{z=y} \ +\ O(g^2)\ ,
\end{aligned}
\end{equation}
The generalized free-action condition in~\eqref{matching01} means that
\begin{equation}
S_0^{\rm b}[T_g^{(0)}\!\phi] \=
-\int \pa_m(T_g\phi)^A\,\pa^m(T_g\phi^*)_A 
\ \buildrel{!}\over{=}\ 
-\int f_g\,\Pi_A^B\,\pa_m\phi^A\,\pa^m\phi^*_B
\= S_g^{\rm b}[\phi] \ ,
\end{equation}
which is met to first order in~$g$ because both sides are equal to
\begin{equation}
-\int \Bigl\{ \pa_m\phi^A\,\pa^m\phi^*_A 
\ -\ \tfrac{g}{\mu^2}\,\phi^*_B\,(\phi^{B}\,\pa_m\phi^{A}+\phi^A\,\pa_m\phi^B)\,\pa^m\phi^*_A
\ +\ O(g^2) \Bigr\} 
\end{equation}
and $\kappa$ cancels out. For the one-loop matching in~\eqref{matching01} we obtain the $O(g)$ contributions
\begin{equation}\label{detma}
\begin{aligned}
S^{\mathrm{b}}_0[T_g\phi]\big|_{O(\hbar)} &\=
\im\tfrac{g}{\mu^2}(N{+}1) \int\Bigl\{
\tfrac12\phi^A\phi^*_A\,\delta(0)\ +\ 
\tfrac{\kappa}{2} \bigl( \phi^A\Box\phi^*_A-\phi^*_A\Box\phi^A\bigr)\,\Box^{-1}(0)
\Bigr\}\ ,
\\[4pt]
- \im\,\tr\ln\sfrac{\delta T_g^{(0)}\!\phi}{\delta\phi} &\= 
\im\tfrac{g}{\mu^2}(N{+}1) \int\Bigl\{
\tfrac12\phi^A\phi^*_A\,\delta(0)\ -\ 
\tfrac{\kappa}{2} \bigl( \phi^A\Box\phi^*_A-\phi^*_A\Box\phi^A\bigr)\,\Box^{-1}(0)
\Bigr\}\ ,
\\[4pt]
S_g^{(1)}[\phi] &\=
\im\tfrac{g}{\mu^2}(N{+}1) \int\Bigl\{
\phi^A\phi^*_A\,\delta(0) \Bigr\}\ ,
\end{aligned}
\end{equation}
which verifies the condition again with the expected cancellation of \(\kappa\).
In \eqref{detma} we use the regularization-dependent quantities 
\(\delta(0)\equiv\delta^{(4)}(x{-}x)\) and \(\Box^{-1}(0)\equiv\Box^{-1}(x{-}x)\), 
which contribute to a mass shift and wavefunction renormalization, respectively.\footnote{
In dimensional regularization both contributions vanish in a massless theory such as the one under consideration. 
With a UV cutoff they are quartically and quadratically divergent, respectively.}
We kept them in this form to maintain the discussion as general as possible and to illustrate 
the matching at first order.

By choosing $\kappa{=}{+}1$ or $\kappa{=}{-}1$ we have the freedom to shift the Nicolai map
entirely to $\phi^A$ or $\phi^*_A$ alone, respectively.
A graphical representation of the Nicolai map to order~$g^2$ looks as follows,
%The following variable rescales the pictures of a factor \scale
\newcommand\scale{.70}
\begin{equation}
\begin{aligned}
T_g\phi\=
\scalebox{\scale}{
\begin{tikzpicture}[baseline={([yshift=-1ex]current bounding box.center)},thick]
\node at (0,0)[circle,fill,inner sep = 2pt] {};
\draw[gluon](1,0)--(0,0);
\end{tikzpicture} 
}
 & \ +\ 
g \
\scalebox{\scale}{
\begin{tikzpicture}[baseline={([yshift=-1ex]current bounding box.center)},thick]
\node at (0,0)[circle,fill,inner sep = 2pt] {};
\draw (1,0)--(0,0);
\draw[gluon] (1+0.75, 0.75)--(1,0);
\draw[gluon] (1+1,0)--(1,0);
\draw[gluon] (1+0.75,-0.75)--(1,0);
\end{tikzpicture} 
}
 \ +\ 
g ^2\
\left(
	\scalebox{\scale}{
	\begin{tikzpicture}[baseline={([yshift=-1ex]current bounding box.center)},thick]
	\node at (-0.25,0)[circle,fill,inner sep = 2pt] {};
	\draw (2,0)--(-0.25,0);
	\draw[gluon] (2+0.75, 0.75)--(2,0);
	\draw[gluon] (2+1,0)--(2,0);
	\draw[gluon] (2+0.75,-0.75)--(2,0);
	\draw[gluon] (1+0.5,-0.9)--(1,0);
	\draw[gluon] (1-0.5,-0.9)--(1,0);
	\phantom{ \draw[gluon] (1-0.5,0.9)--(1,0); }
	\end{tikzpicture} 
	}
	\ +\ 
	\scalebox{\scale}{
	\begin{tikzpicture}[baseline={([yshift=-1ex]current bounding box.center)},thick]
	\node at (0,0)[circle,fill,inner sep = 2pt] {};
	\draw (1,0)--(0,0);
	\draw[gluon] (1+0.75, 0.75)--(1,0);
	\draw[gluon] (1+0.90, 0.40)--(1,0);
	\draw[gluon] (1+1,0)--(1,0);
	\draw[gluon] (1+0.90, -0.40)--(1,0);
	\draw[gluon] (1+0.75,-0.75)--(1,0);
	\end{tikzpicture} 
	}
\right)
\ + \ \cdots
\\
& \ +\ 
\hbar g \
\scalebox{\scale}{
\begin{tikzpicture}[baseline={([yshift=-1ex]current bounding box.center)},thick]
\node at (0,0)[circle,fill,inner sep = 2pt] {};
\draw (0.9,0)--(0,0); 
\draw (0.9,0)--(0.9-  0.75*0.288 , 0.75*1/2); 
\draw (0.9,0)--(0.9+ 0.75*0.288 ,0.75*1/2); 
\draw[gluon] (0.9+0.9,0)--(0.9,0); 
\draw [domain=-30:180+30] plot ( { 0.75* cos(\x)/3 + 0.9  } , {   0.75* sin(\x)/3+0.75* 2/3 } );
\phantom{
\draw [domain=-30:180+30] plot ( { 0.75* cos(\x)/3 + 0.9  } , {-   0.75* sin(\x)/3-0.75* 2/3 } );
}
\end{tikzpicture} 
}
  \ +\ 
\hbar g^2 
\left(
	\scalebox{\scale}{
	\begin{tikzpicture}[baseline={([yshift=-1ex]current bounding box.center)},thick]
	\node at (-1,0)[circle,fill,inner sep = 2pt] {};
	\draw (0.7,0)--(-1,0); 
	\draw (0.7,0)--(0.7-  0.75*0.288 , 0.75*1/2); 
	\draw (0.7,0)--(0.7+ 0.75*0.288 ,0.75*1/2); 
	\draw[gluon] (0.7+0.9,0)--(0.7,0); 
	\draw [domain=-30:180+30] plot ( { 0.75* cos(\x)/3 + 0.7  } , {   0.75* sin(\x)/3+0.75* 2/3 } );
	\phantom{
	\draw [domain=-30:180+30] plot ( { 0.75* cos(\x)/3 + 0.7  } , {-   0.75* sin(\x)/3-0.75* 2/3 } );
	}
	\draw[gluon] (0.5,-0.9)--(0,0);
	\draw[gluon] (-0.5,-0.9)--(0,0);
	\phantom{ \draw[gluon] (1-0.5,0.9)--(1,0); }
	\end{tikzpicture} 
	} 
	 +
	\scalebox{\scale}{
	\begin{tikzpicture}[baseline={([yshift=-1ex]current bounding box.center)},thick]
	\node at (0,0)[circle,fill,inner sep = 2pt] {};
	\draw (0.7,0)--(0,0); 
	\draw (0.7,0)--(0.7-  0.65*0.288 , 0.65*1/2); 
	\draw (0.7,0)--(0.7+ 0.65*0.288 ,0.65*1/2); 
	\draw[gluon] (0.7+0.9,0)--(0.7,0); 
		\draw [domain=-30:180+30] plot ( { 0.65* cos(\x)/3+0.7  } , {   0.65* sin(\x)/3+0.65* 2/3 } );
		\draw[gluon] (0.7+0.4  ,1.1 ) --(0.7,0.65 );
		\draw[gluon] (0.7-0.4  ,1.1  ) --(0.7,0.65 );
	\phantom{
		\draw[gluon] (0.7+0.4  ,-1.1 ) --(0.7,-0.65 );
	}
	\end{tikzpicture} 
	}
		+
	\scalebox{\scale}{
	\begin{tikzpicture}[baseline={([yshift=-1ex]current bounding box.center)},thick]
	\node at (0,0)[circle,fill,inner sep = 2pt] {};
	\draw (1.2,0)--(0,0);
	\draw[gluon] (2.15-0.2, 0.75)--(1.4-0.2,0);
	\draw[gluon] (2.4-0.2,0)--(1.4-0.2,0);
	\draw[gluon] (2.15-0.2,-0.75)--(1.4-0.2,0);
		\draw (0.7,0)--(0.7-  0.75*0.288 , 0.75*1/2); 
		\draw (0.7,0)--(0.7+ 0.75*0.288 ,0.75*1/2); 
		\draw [domain=-30:180+30] plot ( { 0.75* cos(\x)/3+0.7  } , {   0.75* sin(\x)/3+0.75* 2/3 } );
	\end{tikzpicture} 
	}	 
	+ 
	\scalebox{\scale}{
		\begin{tikzpicture}[baseline={([yshift=-1ex]current bounding box.center)},thick]
		\node at (-1,0)[circle,fill,inner sep = 2pt] {};
		\draw (0,0)--(-1,0); 
		\draw (0,0)--( -  0.75*0.288 , 0.75*1/2); 
		\draw (0,0)--( + 0.75*0.288 ,0.75*1/2); 
		\draw[gluon] (0.9,0)--(0,0); 
		\draw [domain=-30:180+30] plot ( { 0.75* cos(\x)/3  } , {   0.75* sin(\x)/3+0.75* 2/3 } );
		\phantom{
		\draw [domain=-30:180+30] plot ( { 0.75* cos(\x)/3  } , {-   0.75* sin(\x)/3-0.75* 2/3 } );
		}
		\draw[gluon] (0.5,-0.9)--(0,0);
		\draw[gluon] (-0.5,-0.9)--(0,0);
		\phantom{ \draw[gluon] (1-0.5,0.9)--(1,0); }
		\end{tikzpicture} 
		} 
\right)
\ +\ \cdots
\\
&  \
\phantom{ \ +\ 
			\hbar g \
			\scalebox{\scale}{
			\begin{tikzpicture}[baseline={([yshift=-1ex]current bounding box.center)},thick]
			\node at (0,0)[circle,fill,inner sep = 2pt] {};
			\draw (1,0)--(0,0); 
			\draw (1,0)--(1-  0.75*0.288 , 0.75*1/2); 
			\draw (1,0)--(1+ 0.75*0.288 ,0.75*1/2); 
			\draw[gluon] (1+0.9,0)--(1,0); 
			\draw [domain=-30:180+30] plot ( { 0.75* cos(\x)/3+1  } , {   0.75* sin(\x)/3+0.75* 2/3 } );
			\phantom{
			\draw [domain=-30:180+30] plot ( { 0.75* cos(\x)/3+1  } , {-   0.75* sin(\x)/3-0.75* 2/3 } );
			}
			\end{tikzpicture} 
			} 
	}
\ +\ 
\hbar^2 g^2 
\left(
	\scalebox{\scale}{
	\begin{tikzpicture}[baseline={([yshift=-1ex]current bounding box.center)},thick]
	\node at (-1,0)[circle,fill,inner sep = 2pt] {};
	\draw (1,0)--(-1,0); 
	\draw (1,0)--(1-  0.75*0.288 , 0.75*1/2); 
	\draw (1,0)--(1+ 0.75*0.288 ,0.75*1/2); 
	\draw[gluon] (1+0.9,0)--(1,0); 
	\draw [domain=-30:180+30] plot ( { 0.75* cos(\x)/3+1  } , {   0.75* sin(\x)/3+0.75* 2/3 } );
	\draw [domain=-30:180+30] plot ( { 0.75* cos(\x)/3   } , {   0.75* sin(\x)/3+0.75* 2/3 } );
	\draw (0,0)--( -  0.75*0.288 , 0.75*1/2); 
	\draw (0,0)--( 0.75*0.288 ,0.75*1/2); 
	\phantom{
	\draw [domain=-30:180+30] plot ( { 0.75* cos(\x)/3+1  } , {-   0.75* sin(\x)/3-0.75* 2/3 } );
	} 
	\end{tikzpicture} 
	} 
	\ +\ 
	\scalebox{\scale}{
	\begin{tikzpicture}[baseline={([yshift=-1ex]current bounding box.center)},thick]
	\node at (0,0)[circle,fill,inner sep = 2pt] {};
	\draw (1,0)--(0,0); 
	\draw (1,0)--(1-  0.65*0.288 , 0.65*1/2); 
	\draw (1,0)--(1+ 0.65*0.288 ,0.65*1/2); 
	\draw[gluon] (1+0.9,0)--(1,0); 
		\draw [domain=-30:180+30] plot ( { 0.65* cos(\x)/3+1  } , {   0.65* sin(\x)/3+0.65* 2/3 } );
		\draw (1,0.65)--(1-  0.65*0.288 , 0.65*1/2 + 0.65 ); 
		\draw (1,0.65)--(1+ 0.65*0.288 , 0.65*1/2 + 0.65 );  
		\draw [domain=-30:180+30] plot ( { 0.65* cos(\x)/3+1  } , {  0.65+ 0.65* sin(\x)/3+0.65* 2/3 } );
	\phantom{
		\draw [domain=-30:180+30] plot ( { 0.65* cos(\x)/3+1  } , {  - 0.65- 0.65* sin(\x)/3-0.65* 2/3 } );
	}
	\end{tikzpicture} 
	}
	\ +\ 
	\scalebox{\scale}{
	\begin{tikzpicture}[baseline={([yshift=-1ex]current bounding box.center)},thick]
	\node at (-1/3,0)[circle,fill,inner sep = 2pt] {};
	\draw (1,0)--(-1/3,0);  
	\draw[gluon] (1+0.9,0)--(1,0); 
		\draw [domain=0:360] plot ( {   (  2 +  cos(\x)  )  /3} , {    sin(\x)  /3  } );
	\end{tikzpicture} 
	}
\right)
\ + \ \cdots
\\
&  \ +\ O( g^3) \ .
\end{aligned}
\end{equation}
Here, the thick dot at the left end of each diagram stands for the argument $x$ of the map,
other vertex positions are integrated over.
Solid lines are free fermion propagators $\slashed\pa\Box^{-1}$ or $\bar{\slashed\pa}\Box^{-1}$,
and wavy lines represent bosonic field insertions $\phi$ or $\phi^*$. 
One of the bosonic legs emanating from each vertex not sourcing a loop carries a derivative (not shown). 
For the full `Nicolai rules', one of course needs to add target-space indices, spinor traces, and weight factors.
All diagrams shown above already appear in the first application of $R_g$ on~$\phi$.
We see that in the $\hbar$ expansion of the map an $r$-loop contribution arises first at order $g^r$, 
so that at each given order in perturbation theory only a finite number of diagrams contribute, as expected.

\section{Adding an auxiliary vector field}\label{s::A}

\noindent
In some field theories with four-fermion interactions one can `resolve' the latter
through a coupling with an auxiliary field~$A$: the Hubbard--Stratonovich transformation.
Schematically, one adds to an action with a $(\bar\psi\psi)^2$ term 
an auxiliary-field coupling $(\bar\psi\psi-A)^2$, schematically
\begin{equation}
\bar\psi\,\im\pa\psi + \tfrac{1}{4}g (\bar\psi\psi)^2 
\qquad\longrightarrow\qquad 
\bar\psi\,\im\pa\psi +  \tfrac{1}{2}g \,A\,\bar\psi\psi - \tfrac{1}{4}gA^2
\= \bar\psi\,\im (\pa-\tfrac{\im}{2} g A)\psi - \tfrac{1}{4}g A^2 \ .
\end{equation}
The four-fermion term has been cancelled, but eliminating~$A$ brings it back.
Hence, the only price is an additional auxiliary field (or several of them). 
Filling in the indices in our schematic argument 
and allowing for the fierzing~\eqref{fierz},
we see that the transformation requires our four-fermion term~\eqref{4fermi} 
to be `factorizable', \ie
\begin{equation}
R_{CD}^{AB} \= \lambda_{\rm s}\, (\bar{m}_i)^{AB}\,(m^i)_{CD}
\ +\ \lambda_{\rm v}\, (\ell_I)_{(C}^ A\,(\ell^I)^B_{D)}
\end{equation}
with some coefficients $\lambda_{\rm s}$ and $\lambda_{\rm v}$
and indices $i$ and~$I$ counting several such terms.
For target geometries with this property we can remove the four-fermion interaction
by introducing a bunch of complex scalar and real vector auxiliary fields
\begin{equation}
A^i \= (m^i)_{CD}\,\psi^C \psi^D \quad\und\quad
A^I_m \= (\ell^I)^B_D\,\psi^D \sigma_m \bar\psi_B\ ,
\end{equation}
respectively.

This is actually the case for all hermitian symmetric spaces~\cite{HN1,HN2}.\footnote{
See~\cite{review} for a more general review on nonlinear realization
and hidden local symmetries.}
Therefore, these geometries allow for an auxiliary-field reformulation.
The complex projective spaces~$\C{\rm P}^N$ treated above are the maximally
symmetric compact examples, and indeed \eqref{Rfact} shows that 
a single real vector auxiliary~$A_m$ suffices. 
Actually, the Hubbard--Stratonovich trick can be slightly generalized by shifting~$A$
by an arbitrary function of bosonic fields. We make use of this option and choose
\begin{equation}\label{Aval}
A_m \= \tfrac{1}{\mu^2}\,f_g\,\bigl[
\im\,\phi^C \pa_m\phi^*_C - \im\,\phi^*_C\,\pa_m\phi^C
+ \psi^A\, \Pi_A^B\, \sigma_m\,\bar\psi_B \bigr]\ ,
\end{equation}
where the form of the (arbitrary) bosonic contribution will be justified later on.
With some algebra this yields the enhanced lagrangian
\begin{equation}\label{cae}
\begin{aligned}
\widetilde\Lagr &\= 
-f_g\,\Pi^B_A \pa_m\phi^A\pa^m\phi^*_B 
+\tfrac14 f_g^2\tfrac{g}{\mu^2}(\phi^A\pa_m\phi^*_A-\phi^*_A\pa_m\phi^A)^2
+\tfrac{\im}{2} f_g\,g\,(\phi^A\pa_m\phi^*_A-\phi^*_A\pa_m\phi^A)\,A^m
-\tfrac14 g\mu^2 A_m A^m
\\[4pt] & \qquad\!
- \tfrac{\im}{2}  f_g\, \Pi^{ B}_A
( \psi^A \slashed{\pa} {\bar\psi}_B + {\bar\psi}_B \bar{\slashed{\pa}} \psi^A)
+ \tfrac12 f_g\,g\,\Pi^B_A\,\psi^A\slashed{A}\,\bar\psi_B
+ \tfrac{\im}{2} f^2_g \tfrac{g}{\mu^2} \, \Pi^B_A 
( \psi^A \phi^C \slashed{\pa} \phi^*_B {\bar \psi}_C
+ \bar\psi_B \phi^*_C \bar{\slashed{\pa}}\phi^A \psi^C )
\ .
\end{aligned}
\end{equation}

There exists an instructive superfield formulation~\cite{HN1,HN2},
\begin{equation} \label{gaction}
\widetilde{S}_{\textrm{\tiny SUSY}} \ = \int\!\diff^4x \ \widetilde\Lagr \qquad\textrm{with}\qquad
\widetilde\Lagr \= \int\!\diff^2\theta\,\diff^2\bar\theta \;\bigl\{ \ep^{g V}\;\Phi_a^\dagger\Phi^a\;-\;\mu^2\,V \bigr\}\ ,
\end{equation}
and corresponding expressions for $\widetilde\Delta$ and~$\widetilde\Delt$,
where the auxiliary real vector superfield~$V$ with components~$(C,L,A_m,\lambda,\chi,D)$ 
expands in $x$~coordinates as
\begin{equation}
V \= -C -\im\theta\chi +\im\bar\theta\bar\chi -\tfrac{\im}{2}\theta^2 L +\tfrac{\im}{2}\bar\theta^2 L^*
+ \theta\sigma^m\bar\theta A_m 
-\im\theta^2\bar\theta\bigl[\bar\lambda+\sfrac{\im}{2}\bar{\slashed{\partial}}\chi\bigr] 
+\im \bar\theta^2\theta\bigl[\lambda+\sfrac{\im}{2}\slashed{\partial}\bar\chi\bigr]
-\sfrac12\theta^2\bar\theta^2\bigl[D+\sfrac12 \Box C\bigr]\ .
\end{equation}
The action~\eqref{gaction} enjoys a complexified local U(1) invariance under
\begin{equation}\label{gatr}
V\ \mapsto\ V\,+\,\im\Lambda\,-\,\im\Lambda^\dagger \qquad\und\qquad
\Phi^a\ \mapsto\ \ep^{-\im g\Lambda}\,\Phi^a
\end{equation}
with a chiral superfield parameter~$\Lambda$. 
The coordinate choice~\eqref{gauge} is not compatible with the Wess--Zumino gauge,
and it completely breaks the gauge symmetry.
Indeed, eliminating the vector superfield by its algebraic equation of motion,
\begin{equation}
V \= -\frac1g\,\log \Bigl[ \frac{g}{\mu^2} \Phi_a^\dagger\Phi^a \Bigr] \= - \frac{1}{\mu^2} {\rm K}[\Phi^a,\Phi_a^\dagger]\ ,
\end{equation}
brings back the original action \eqref{Kaction} with the $\C{\rm P}^N$ K\"ahler potential~\eqref{Kcpn}
(up to a constant). We can interpret the action \eqref{gaction} and the gauge transformation \eqref{gatr} 
as a way of performing the $\C{\rm P}^N$  identification of scalar superfields under complex chiral parameters: 
the supergauge transformation realizes in a supersymmetric fashion  a  
\(\GroupName{U}(1)\)  local transformation, thus extending the standard gauge invariance 
to a \(\GroupName{U}(1)_\C\) symmetry including conformal rescaling. 

Let us go now to the component level and eliminate auxiliary fields $F$, $L$, $\chi$, $\lambda$, $C$ and $D$ 
from the action by their equations of motion, \eg
\begin{equation}
F^a \= -\psi^a\,\tfrac{\psi^b \phi_b^*}{\phi^*\phi} \quad,\qquad
\chi = -\tfrac{\im\sqrt{2}}{g}\,\tfrac{\psi^a\phi^*_a}{\phi^*\phi} \quad,\qquad
g\,\ep^{-g C}\,\phi^*\phi \= \mu^2\ .
\end{equation}
After lengthy but straightforward computations we arrive at
\begin{equation}
\begin{aligned}\label{Ltilde}
\widetilde\Lagr&\= f_g\, \Pi^b_a \Bigl\{ 
-D_m\phi^a D^m\phi^*_b
-\tfrac{\im}{2}\bigl[\psi^a\slashed{D}\bar\psi_b+\bar\psi_b\bar{\slashed{D}}\psi^a\bigr]
+\tfrac{\im}{2}\tfrac{1}{\phi^*\phi}\bigl[
\psi^a \phi^c \slashed{D} \phi^*_b \bar\psi_c
+\bar\psi^b \phi^*_c \bar{\slashed{D}} \phi^a \psi^c \bigr]\Bigr\} \\[4pt]
& \quad + \tfrac14 f_g^2 \tfrac{g}{\mu^2}\,\bigl(\phi^a D_m\phi^*_a - \phi^*_a D_m\phi^a)^2
\end{aligned}
\end{equation}
where $D_m=\pa_m-\tfrac{\im}2\, g\,A_m$ is a U(1)-covariant derivative.
We note that $\Pi^b_aD_m\phi^a=\Pi^b_a\pa_m\phi^a$.
For the penultimate components we find
\begin{equation}
\widetilde\Delt \= -\im\sqrt{2}\,f_g\,\Pi_a^b\,\slashed{D}\phi^a\,\bar\psi_b
+ \im\sqrt{2} f_g^2\tfrac{g}{\mu^2}\,
(\phi^b\slashed{D}\phi^*_b-\phi^*_b\slashed{D}\phi^b)\,\phi^a\bar\psi_a\ .
\end{equation}
Both $\widetilde\Lagr$ and $\widetilde\Delt$ are manifestly gauge invariant.
We are left with the fields $\phi^a$, $\psi^a$ and~$A_m$. 
Eliminating the latter via its equation of motion
\begin{equation} \label{Aeom}
\mu^2 A_m\= f_g \bigl[ \im\phi^c\pa_m\phi_c^*-\im\phi_c^*\pa_m\phi^c
+ \Pi^b_a\,\psi^a\sigma_m\bar\psi_b \bigr] \qquad\Leftrightarrow\qquad
\im\phi^c D_m\phi_c^* - \im\phi_c^* D_m\phi^c \= - \Pi_a^b\,\psi^a \sigma_m\bar\psi_b
\end{equation}
reverts to the four-fermion interaction, so we keep \(A_m\) in the lagrangian.  

Instead, we now employ the local supersymmetric U$(1)_\C$ invariance to fix one of the chiral superfields,
$\Phi^0 =\sqrt{\tfrac{\mu^2}{g}}$, which explicitly connects the auxiliary-superfield formulation 
with our coordinate choice~\eqref{gauge} for the nonlinear sigma model.
It is not hard to see that this gauge fixing comes with a trivial Faddeev--Popov determinant.
The gauge-fixed lagrangian then indeed agrees with~\eqref{cae}.
Using the identities 
$\Pi^B_A\phi^A=f_g\phi^B$ and $\tfrac{g}{\mu^2}\phi^A\phi^*_A=f_g^{-1}-1$ 
as well as
\begin{equation}
\begin{aligned}
&\phi^c D\phi^*_c-\phi^*_c D\phi^c \=
\phi^C \pa\phi^*_C-\phi^*_C \pa\phi^C + \im f_g^{-1} \mu^2 A \ ,
\\[4pt]
&\Pi^b_a D \phi^a\bar\psi_b \= 
\Pi^B_A D\phi^A \bar\psi_B + \tfrac{\im}{2} f_g\,g\,A\,\phi^A\bar\psi_A\ ,
\end{aligned}
\end{equation}
we also obtain
\begin{equation} \label{tildedelton}
\widetilde\Delt \= -\im\sqrt{2}f_g\,\Pi_A^B\,\slashed{\pa}\phi^A\,\bar\psi_B
+ \im\sqrt{2} f_g^2\tfrac{g}{\mu^2}\,(\phi^C\slashed{\pa}\phi^*_C-\phi^*_C\slashed{\pa}\phi^C)\,\phi^A\bar\psi_A
-\sqrt{2} f_g\, g\,\slashed{A}\,\phi^A\bar\psi_A\ .
\end{equation}
Furthermore, after this gauge-fixing the value of \(A_m\) in   \eqref{Aeom} is exactly \eqref{Aval}.

Again, we can set up the Nicolai map. Employing
\begin{equation}
\delta_\alpha Y\=\sqrt{2}\,{\psi}_\alpha\,\tfrac{\delta Y}{\delta{\phi}}
\ +\ \delta^\alpha A_m\,\tfrac{\delta Y}{\delta A_m}\ ,
\end{equation}
with a rather complicated expression for $\delta^\alpha A_m$, which we shall not reproduce here, and~\footnote{
Again, we note that this is not $\pa_g\eqref{tildedelton}$ but the $g$-derivative is taken 
before eliminating any auxiliary, cf.~footnote~7.}
\begin{equation}
\tfrac{\mu^2}{\sqrt{2}}\,\partial_g\widetilde\Delt_\alpha\=
\im\,\phi^B\,\slashed{\pa}_{\alpha\dot\alpha} (\phi^*_B \phi^A)\,\bar{\psi}^{\dot\alpha}
- \mu^2\slashed{A}_{\alpha\dot\alpha}\phi^A\bar{\psi}^{\dot\alpha}
- \phi^A\psi^B_\alpha(\bar\psi_A\bar\psi_B) 
+\ O(g) \ ,
\end{equation}
we may compose the coupling flow operator
\begin{equation}
\widetilde{R}_g \= -\tfrac{\im}{4} \tfrac{1+\kappa}{2} \int\!\diff{y}\ \pa_g
\bcontraction{}{\widetilde\Delt}{(y)_\alpha\ }{\delta}
\widetilde\Delt(y)_\alpha\ \delta^\alpha
\ -\ \tfrac{\im}{4} \tfrac{1-\kappa}{2} \int\!\diff{y}\ \pa_g
\bcontraction{}{\bar{\widetilde\Delt}}{(y)^{\dot\alpha}\ }{\bar\delta}
\bar{\widetilde\Delt}(y)^{\dot\alpha}\ \bar\delta_{\dot\alpha}\ ,
\end{equation}
with the contractions now being defined in a $(\phi,A)$ background.\footnote{
We have omitted auxiliary-field contractions coming from terms nonlinear in auxiliary fields,
as we do not need them here.} 
Even though the fermions appear just quadratically in the lagrangian~\eqref{Ltilde}, we
do not arrive at a purely classical Nicolai map, in the sense described in the Introduction.
To leading order in the coupling~$g$, its classical part looks as follows,
\begin{equation}
(T_g^{(0)}\!\phi)^A(x) \= {\phi}^A(x)\  +\ \tfrac{1+\kappa}{2}\tfrac{g}{\mu^2}\!\int\!\diff{y}\ 
\bigl[ \phi^B\, 
\pa_m(\phi^*_B\phi^A) 
+ \im\mu^2 A_m \phi^A\bigr](y)\,
\pa^m\Box^{-1}(y{-}x) +\ O(g^2)\ ,
\end{equation}
and similarly for $(T_g^{(0)}\!A)_m$.
Checking the free-action condition
\begin{equation} \label{clash1}
\begin{aligned}
\int \pa_m(T_g^{(0)}\!\phi)^A\,\pa^m(T_g^{(0)}\!\phi^*)_A 
\ &\buildrel{!}\over{=}\
\int \bigl\{ f_g\Pi^B_A\,\pa_m\phi^A\,\pa^m\phi^*_B
-\tfrac14 f_g^2 \tfrac{g}{\mu^2}(\phi^C\pa_m\phi^*_C {-} \phi^*_C\pa_m\phi^C)^2 \\
&\qquad -\tfrac{\im}{2} f_g\,g\,(\phi^C\pa_m\phi^*_C {-} \phi^*_C\pa_m\phi^C)\,A^m
+\tfrac14 g\mu^2 A_m A^m \bigr\}
\end{aligned}
\end{equation}
we obtain the first-order requirement
\begin{equation} \label{clash2}
\begin{aligned}
\int\Bigl\{ 
(\phi^*\phi)(\pa\phi^*{\cdot}\pa\phi) + 
\tfrac{1+\kappa}{2} (\phi\,\pa\phi^*){\cdot}(\phi\,\pa\phi^*) +
\tfrac{1-\kappa}{2} (\phi^*\pa\phi){\cdot}(\phi^*\pa\phi) +
\tfrac{1+\kappa}{2} \im\mu^2 A{\cdot}(\phi\,\pa\phi^*) -
\tfrac{1-\kappa}{2} \im\mu^2 A{\cdot}(\phi^*\pa\phi) \Bigr\} \\
\buildrel{!}\over{=}\
\int\Bigl\{
(\phi^*\phi)(\pa\phi^*{\cdot}\pa\phi) + 
(\phi^*\pa\phi){\cdot}(\phi\,\pa\phi^*) -
\tfrac14 \bigl[\mu^2 A - \im(\phi\,\pa\phi^*{-}\phi^*\pa\phi)\bigr]{\cdot}
\bigl[\mu^2 A - \im(\phi\,\pa\phi^*{-}\phi^*\pa\phi)\bigr] \Bigr\} \ ,
\end{aligned}
\end{equation}
where the dots indicate Lorentz contractions, 
and the round brackets enclose target coordinate contractions.
We observe that the left-hand side is linear in~$A$ while the right-hand side is quadratic.
Matching both sides (and cancelling $\kappa$) requires putting
\begin{equation}
\mu^2 A_m \ \buildrel{!}\over{=}\
\im\,\bigl( \phi^C \pa_m \phi^*_C - \phi^*_C \pa_m \phi^C \bigr) \ +\ O(g) \ +\ O(\hbar)\ ,
\end{equation}
which agrees with the equation of motion~\eqref{Aeom} after integrating out the fermions!
The crux of the mismatch, however, lies in the absence of~$A$ in the free action,
which renders the $g\to0$ limit singular for the auxiliary field.
In other words, the propagator for $A$ is proportional to $\tfrac{1}{g}$, 
which upsets the perturbative expansion in~$g$ (not in powers of the fields)
of correlation functions, with or without the Nicolai map.
The remedy is to integrate out the auxiliary vector~$A$ and work with an effective theory of~$\phi$ alone.
This, however, revives the four-fermion interaction:
the classical solution~\eqref{Aeom} shows that $A$ yields a fermion loop, and the ultralocal propagator
$\langle A(x)\,A(y)\rangle\sim\delta(x{-}y)$ glues two such loops together, 
effectively reproducing the four-fermion interaction.
It seems that the auxiliary-field reformulation of the supersymmetric nonlinear sigma models 
does not simplify the Nicolai map in the end. 
However, the situation is somewhat reminiscent of the super Yang--Mills case \cite{LR1}, 
where one needs to use specific field redefinitions to define the map. 
A similar construction might apply to this case too and deserves further study.

Let us conclude with a comment on   the \(N{=}1\) case.  
We have the accidental isomorphism \(\C \mathrm P^1 \simeq S^2\), the real 2-sphere, 
which is maximally symmetric in the real sense and whose Riemann tensor admits therefore 
the standard decomposition \(\mathrm{Riem} = gg -gg \) in real coordinates. 
In this case indices \(A,B,\dots\) have only one value and \(\Pi \equiv f_g\). 
Correspondingly, the four-fermion term of the lagrangian~\eqref{LagrCPN} takes the form
\begin{equation}
\Lagr_{\psi^4} \= -\tfrac12 f_g^3 \tfrac g {\mu^2} \, \psi \psi \, \bar \psi \bar \psi\ ,
\end{equation}
which can be resolved by means of a standard Hubbard--Stratonovich auxiliary complex scalar \( A \sim  \psi \psi \). 
Of course, the clash with the perturbative expansion in \(g\) described in~\eqref{clash1}-\eqref{clash2} 
applies also in this case. 

\subsection*{Acknowledgments}
We thank Hermann Nicolai for discussions.

\newpage

%\section*{}

\end{document}